\documentclass[conference]{IEEEtran}
\IEEEoverridecommandlockouts
\usepackage{cite}
\usepackage{amsmath,amssymb,amsfonts}
\usepackage{comment}
\usepackage{algorithmic}
\usepackage[dvipdfmx]{graphicx}
\usepackage{graphicx}
\usepackage{textcomp}
\usepackage{xcolor}
\usepackage{empheq}
\usepackage{flushend}
\usepackage{fancyhdr}
\def\BibTeX{{\rm B\kern-.05em{\sc i\kern-.025em b}\kern-.08em
    T\kern-.1667em\lower.7ex\hbox{E}\kern-.125emX}}
\begin{document}

\title{Spatially Continuous Non-Contact Cold Sensation Presentation Based on Low-Temperature Airflows
\thanks{This work was partly supported by grants from JSPS KAKENHI (JP21H03474, JP21K19778), and funded by Tohin Co,.Ltd.}
}

\author{
\IEEEauthorblockN{Koyo~Makino}
\IEEEauthorblockA{\textit{Grad. Sch. of Science and Technology} \\
\textit{University of Tsukuba}\\
Ibaraki, Tsukuba
}
\and
\IEEEauthorblockN{Jiayi~Xu}
\IEEEauthorblockA{\textit{Grad. Sch. of Science and Technology} \\
\textit{University of Tsukuba}\\
Ibaraki, Tsukuba
}
\and
\IEEEauthorblockN{Akiko Kaneko}
\IEEEauthorblockA{\textit{Fac. of Engineering, Information and Systems} \\
\textit{University of Tsukuba}\\
Ibaraki, Tsukuba
}
\and
\IEEEauthorblockN{\hspace{12mm}
Naoto~Ienaga}
\IEEEauthorblockA{\textit{\hspace{12mm}
Fac. of Engineering, Information and Systems} \\
\textit{\hspace{12mm}
University of Tsukuba}\\
\hspace{12mm}
Ibaraki, Tsukuba
}
\and
\IEEEauthorblockN{Yoshihiro~Kuroda}
\IEEEauthorblockA{\textit{Fac. of Engineering, Information and Systems} \\
\textit{University of Tsukuba}\\
Ibaraki, Tsukuba
}
}

\maketitle
\thispagestyle{fancy}
\begin{abstract}
Our perception of cold enriches our understanding of the world and allows us to interact with it. Therefore, the presentation of cold sensations will be beneficial in improving the sense of immersion and presence in virtual reality and the metaverse. This study proposed a novel method for spatially continuous cold sensation presentation based on low-temperature airflows. We defined the shortest distance between two airflows perceived as different cold stimuli as a local cold stimulus group discrimination threshold (LCSGDT). By setting the distance between airflows within the LCSGDT, spatially continuous cold sensations can be achieved with an optimal number of cold airflows. We hypothesized that the LCSGDTs are related to the heat-transfer capability of airflows and developed a model to relate them. We investigated the LCSGDTs at a flow rate of 25~L/min and presentation distances ranging from 10 to 50 mm. The results showed that under these conditions, the LCSGDTs are 131.4 ± 1.9 mm, and the heat-transfer capacity of the airflow corresponding to these LCSGDTs is an almost constant value, that is, 0.92. 
\end{abstract}


\section{INTRODUCTION}
\par Cold sensation is one of the most significant somatic sensations since it enhances our perception of the world and allows us to interact with world even without direct contact\cite{warmAndCool}. For instance, we experience cold when putting hands in a river or when walking on a snow-covered mountain. In addition, this kind of coldness is not limited to a single part of the body, but extends to the whole body in a spatially continuous manner. The presentation of spatially continuous cold sensations will result in a revolutionary type of interaction and will be used in the field of virtual reality (VR) and the Metaverse to enable users to immerse themselves more fully in simulated virtual environments.
\begin{figure}[tb]
\begin{center}
\includegraphics[scale=0.27]{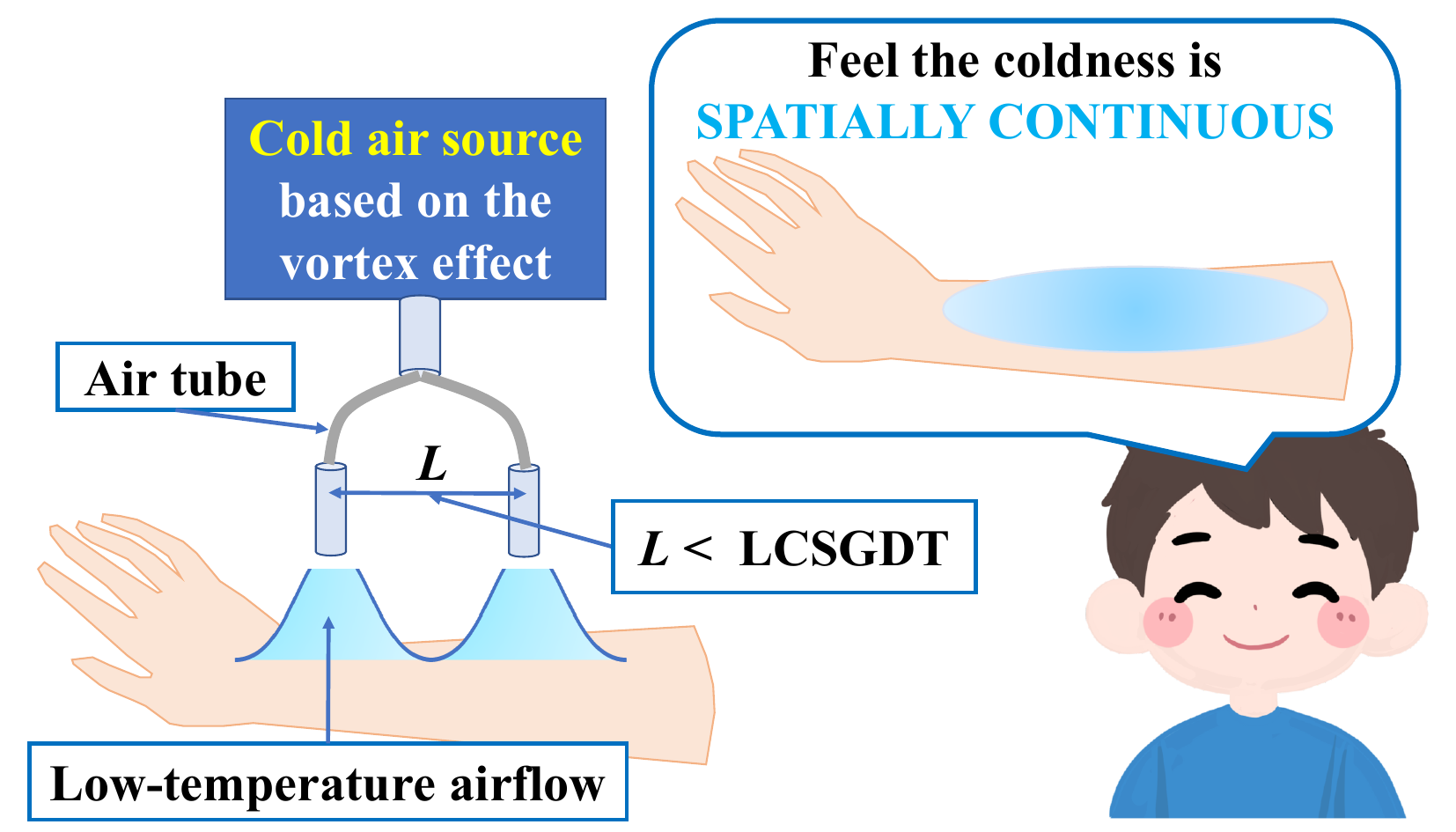}
\end{center}
\vspace{-3mm}
\caption{Spatially continuous non-contact cold sensation presentation based on low-temperature air flow. By placing the distance between the airflows within the local cold stimulus group discrimination threshold (LCSGDT), spatially continuous cold sensations can be achieved.}
\label{fig:sep}
\end{figure}
\par For spatially continuous cold sensation presentation, cold stimuli need to be applied to a wide area of the skin. Additionally, the cold stimuli should be sensed as a whole rather than as separate points. If the minimum distance between cold stimuli that are perceived as separate points can be 
determined, it is possible to achieve spatially continuous cold presentation with a small number of cold stimuli by placing the stimuli within this distance. We refer to this distance as the local cold stimulus group discrimination threshold (LCSGDT).
\par Among the various methods of delivering cold stimuli, non-contact methods have gained considerable attention due to their ability to eliminate the discomfort caused by mechanical contact. Some studies proposed a method that utilizes ultrasound to deliver mist water to the skin for providing cold stimuli ~\cite{m0,m1}. The aim of this method is to achieve cold sensations that can be pinpointed and located. Another study proposed a method for providing cold stimuli through the use of low-temperature airflow generated by the vortex effect~\cite{a0,a1}. The cold stimulus provided by this method can be spatially diffused in a wide area of skin. To this end, we believe that by using a group of cold stimuli generated by low-temperature airflows, it will be possible to achieve spatially continuous cold sensations.
\par In this study, we propose a method for eliciting spatially continuous cold sensations based on low-temperature airflows by placing the distance between the airflows within the LCSGDT, as shown in Fig.~\ref{fig:sep}. We hypothesize that the LCSGDT depends on the heat-transfer capability of airflows. To this end, we introduce an airflow heat-transfer model, and investigate how it relates to the threshold. This paper describes the prototype system and the evaluation results of the proposed method and the hypothesis.

\section{RELATED WORKS}
\par Most conventional studies used Peltier devices to provide cold stimuli. For example, Peiris et al.~\cite{p0} developed a head-mounted display with Peltier devices to display cold stimuli suitable for the virtual world. Ito et al.~\cite{p1} proposed using Peltier devices to provide cold stimuli to simulate a virtual underwater experience. Additionally, G\"{u}nther et al.~\cite{w1} proposed a method that utilized flexible tubes filled with water to deliver cold stimuli. The use of these methods requires mechanical contact with the skin, but prolonged close contact may cause discomfort.
\par As for providing cold stimuli without mechanical contact, air conditioners are the traditional method~\cite{first}~\cite{AoEs}. However, this method cannot produce cold stimuli immediately because it requires replacing the entire air environment around the user. In regards to providing cold stimuli immediately, Nakajima et al.~\cite{m0,m1} proposed a method that uses focused ultrasound to accelerate the vaporization of the mist delivered to human skin. This method is capable of pinpointing cold stimuli. On the other hand, Xu et al.~\cite{a0,a1} proposed a method using low-temperature airflow to generate cold stimuli. This method is capable of immediately providing spatially diffused cold stimuli over a wide area of skin. 
\par In this study, we propose using low-temperature airflows to provide spatially continuous cold sensations. When the distance between airflows is within the minimum distance between cold stimuli perceived as separate points, spatially continuous cold sensations can be achieved.

\section{METHOD}
\par In this section, we describe the proposed method to elicit spatially continuous cold sensations and the hypothesis for determining the LCSGDT. In addition, we introduce a model that describes the heat-transfer capability of low-temperature airflow since we assume that the LCSGDT is related to it.

\subsection{Hypothesis of local cold stimulus group discrimination threshold}
\par
\begin{figure}[t]
\begin{center}
\includegraphics[scale=0.30]{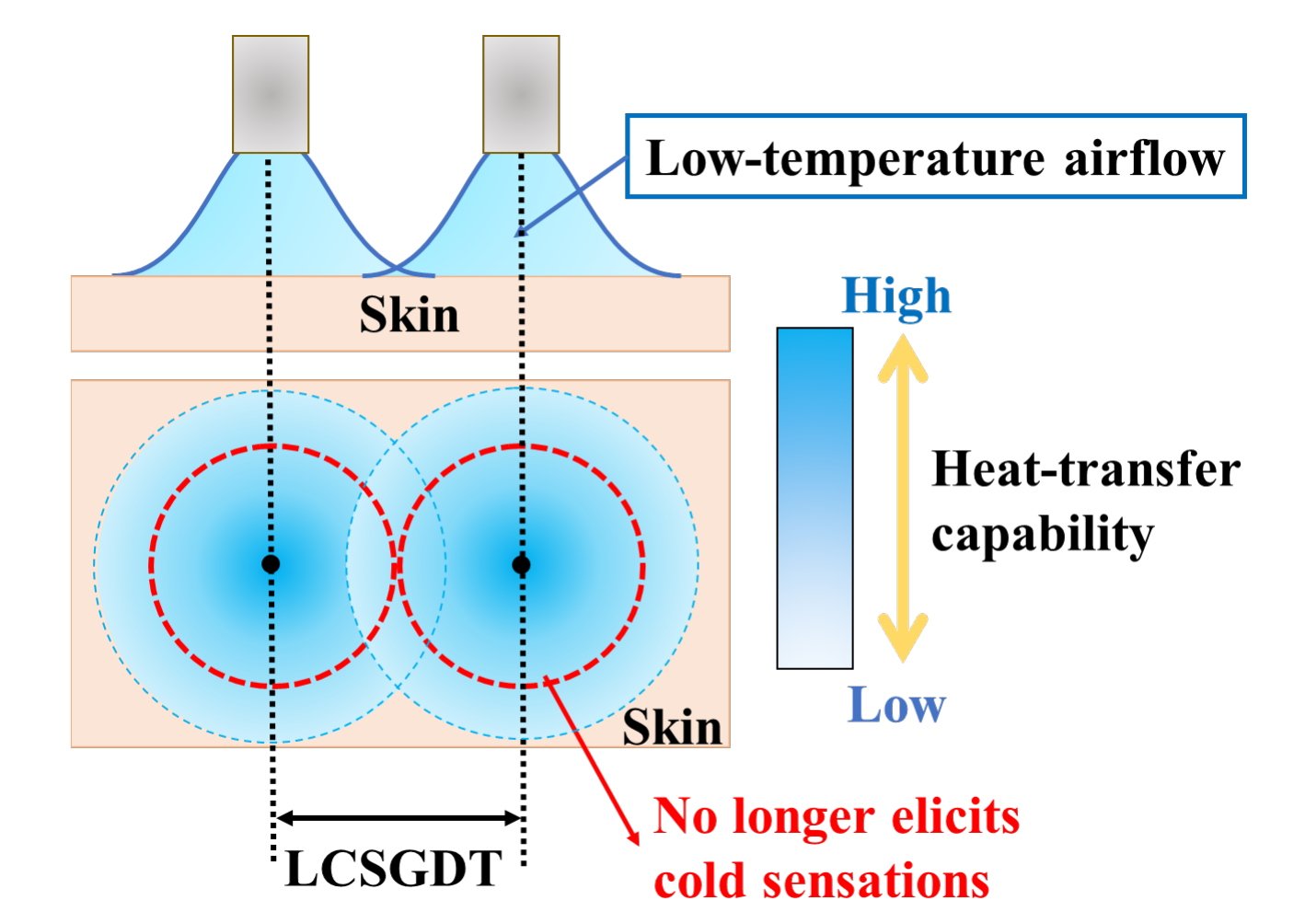}
\end{center}
\vspace*{-3mm}
\caption{Hypothesis of local cold stimulus group discrimination threshold.}
\label{fig:AirflowHeatTransfer}
\end{figure}

 The vortex effect is capable of producing airflow at low temperatures (e.g., 0~$^\circ$C) instantly from a compressed air supply~\cite{vt}. 
As shown in Fig.~\ref{fig:AirflowHeatTransfer}, as airflows impinge on the skin surface, they gradually spread out in a circular pattern from the center, with the largest heat-transfer capability occurring at the center and decreasing radially outward from there. Once the heat-transfer capability decreases to a certain level, the resulting temperature change no longer elicits cold sensations. We hypothesize that superimposing airflows at a distance that is slightly shorter than that at which the cold sensation becomes imperceptible can achieve spatially continuous cold sensations. In other words, twice the distance at which the cold sensation becomes imperceptible is the LCSGDT.

\subsection{Airflow heat-transfer model}
\par
\begin{figure}[t]
\begin{center}
\includegraphics[scale=0.3]{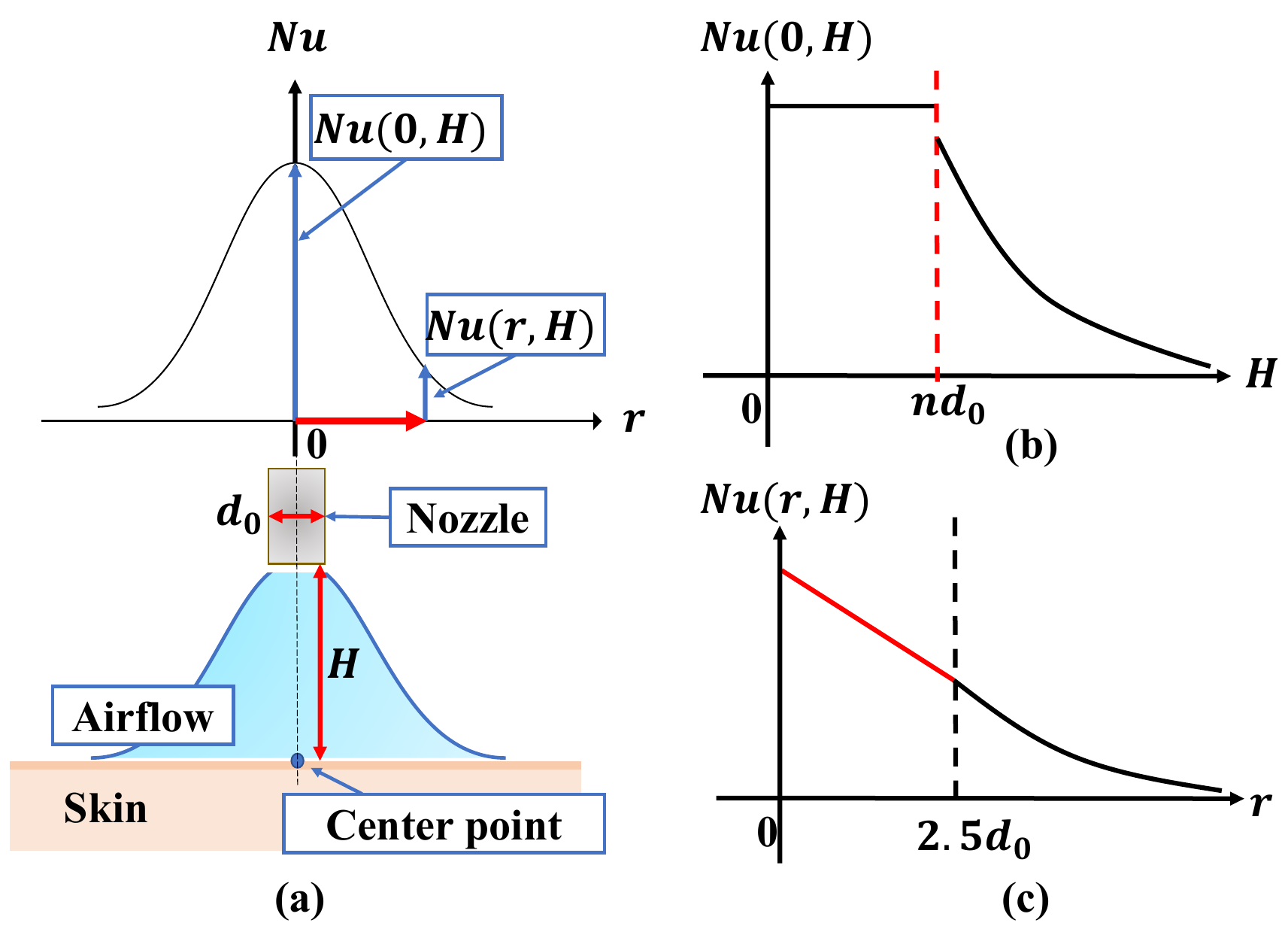}
\end{center}
\vspace*{-3mm}
\caption{Airflow heat-transfer model with diagram of the cooling process (a). The Nusselt number $Nu(0,H)$ of the center point according to the presentation distance $H$ from nozzle to skin (b). The Nusselt number $Nu(r,H)$ according to the diffusion radius $r(r>0)$ (c).}
\label{fig:as1}
\end{figure}

 To investigate how the heat-transfer capability of low-temperature airflow relates to the LCSGDT, we introduce an airflow heat-transfer model. In jet flow engineering, a dimensionless number named the Nusselt number is used to represent the heat-transfer capability of jets such as airflows and waterflows~\cite{Nu}. As shown in the Fig.~\ref{fig:as1}~(a), the Nusselt number $Nu(r,H)$ changes along the center point to the diffusion radius $r$. The empirical formula for the Nusselt number $Nu(0,H)$ of the center point is given by
\begin{empheq}[left={Nu(0, H)=\empheqlbrace}]{alignat=2}
&0.94Pr^{0.4}Re^{0.5} & \quad (&H\leq 4d_0) \label{eq:Nu01}\\
&11.6Pr^{0.5}Re^{0.5}\left(\dfrac{d_0}{H}\right) & \quad (&H\geq 10d_0),\label{eq:Nu02}
\end{empheq}
where $H$ is the presentation distance from the nozzle, $d_{0}$ is the nozzle inner diameter, $Pr$ is the Prandtl number, and $Re$ is the Reynolds number. It should be noted that the Reynolds number is dependent on the volume flow rate of the airflow.
The empirical formula for the Nusselt number $Nu(r,H)$ of the diffusion radius $r(r>0)$ is given by 
\begin{align}
  Nu(r, H) &= G\left(r, H\right)F\left(Re\right)Pr^{0.42}, ~\label{eq:Nu(r, H)}\\
  G\left(r, H\right)&= \dfrac{d_0}{r}\dfrac{1-1.1\dfrac{d_0}{r}}{1+0.1\left(\dfrac{H}{d_0}-6\right)\dfrac{d_0}{r}},\\
  F(Re) &=  2Re^{0.5}\left(1+\dfrac{Re^{0.55}}{200}\right)^{0.5},
\end{align}
where $2.5d_0\leq r\leq 7.5d_0$, $2\times10^{3}\leq Re\leq 4\times10^{5}$, $2d_0<H<12d_0$.

These empirical formulas are based on experiments performed with waterflows as jets~\cite{a}. Our study aims to develop a heat-transfer model for low-temperature airflows. Based upon the above-mentioned empirical formulas, we propose the following assumptions to improve these formulas so that they may be applied to our method:
\begin{enumerate}
\item The conventional empirical formulas have not defined the Nusselt number $Nu(0, H)$ of the center point at all presentation distances $H$. As shown in Fig.~\ref{fig:as1}~(b), we consider that until the presentation distance reaches a certain value $n d_0$, the Nusselt number $Nu(0, H)$ is maintained at a certain value. Following that, the Nusselt number $Nu(0, H)$ decreases with the presentation distance $H$. In other words, when the presentation distance $H$ is less than $n d_0$,  we calculated the Nusselt number $Nu(0, H)$ based on (\ref{eq:Nu01}). When the presentation distance $H$ is greater or equal to $n d_0$,  we calculated the Nusselt number $Nu(0, H)$ based on (\ref{eq:Nu02}).
\item The conventional empirical formulas have not defined the Nusselt number $Nu(r, H)$ at all diffusion radius $r$. As shown in Fig.~\ref{fig:as1}~(c), when the diffusion radius $r$ is less than $2.5 d_{\rm 0}$, we consider the Nusselt number $Nu(r, H)$ decreases with diffusion radius $r$ linearly for simplicity. 
When the diffusion radius $r$ is greater or equal to $2.5 d_{\rm 0}$, the Nusselt number $Nu(r,H)$ changes based on (\ref{eq:Nu(r, H)}).
\item We consider it necessary to re-estimate some coefficients to adapt the conventional empirical formulas to low-temperature airflow.
\end{enumerate}
Based on these assumptions, we construct the airflow heat-transfer model. Specifically, the Nusselt number $Nu(0,H)$ of the center point is given by
\begin{empheq}[left={Nu(0, H)=\empheqlbrace}]{alignat=2}
&\beta Pr^{0.4}Re^{0.5} & \quad (&H < n d_0) \\
&\gamma Pr^{0.5}Re^{0.5}\left(\dfrac{d_0}{H}\right) & \quad (&H\geq n d_0),
\end{empheq}
where $\beta$, and $\gamma$ are the coefficients that need to be reestimated. 
The Nusselt number $Nu(r,H)$ of the diffusion radius $r(r>0)$ is given by 
\begin{empheq}[left={Nu(r, H)=\empheqlbrace}]{alignat=2}
&E(r, H)r + Nu(0,H) & \quad (&r < 2.5 d_0) \\
&G\left(r, H\right)F\left(Re\right)Pr^{0.42} & \quad (&r\geq 2.5d_0),
\end{empheq}
\vspace{2mm}
\begin{align}
  E(r, H) &= \dfrac{Nu(2.5d_0, H)-Nu(0, H)}{2.5d_0},\\
  G\left(r, H\right)&= \dfrac{d_0}{r}\dfrac{1-a\dfrac{d_0}{r}}{1+b\left(\dfrac{H}{d_0}-c\right)\dfrac{d_0}{r}},\\
  F(Re) &=  2Re^{0.5}\left(1+\dfrac{Re^{0.55}}{f}\right)^{0.5},
\end{align}
where $n$, $a$, $b$, $c$, and $f$ are the coefficients that needs to be reestimated.

\subsection{Relationship between temperature change and heat-transfer model}
\par In our airflow heat-transfer model, coefficients $\beta$, $\gamma$, $n$, $a$, $b$, $c$, and $f$ are unknown. We aim to estimate these unknown coefficients by measuring the temperature change of an object caused by low-temperature airflow impinging on its surface. To this end, we investigate the relationship between temperature change and the Nusselt number. When airflow impinges on the object's surface, the heat $q(r, H)$ transferred per unit area from the airflow to the object by convective heat-transfer can be calculated as
\begin{equation}
q(r, H)= h_{\rm s}(r, H)(T_{\rm s0} - T_{\rm a}(H)),
\end{equation}
where $h_{\rm s}(r, H)$ is the heat-transfer coefficient, $T_{\rm s0}$ is the initial temperature of object's surface, and $T_{\rm a}(H)$ is temperature of the airflow when it impacts the skin. Among them, the heat-transfer coefficient $h_{\rm s}(r, H)$ is proportional to the Nusselt number $Nu(r, H)$~\cite{Nu}, and
 the heat $q(r, H)$ is proportional to the temperature change $\Delta T_{\rm s}(r, H)$ of object's surface. 
\begin{eqnarray}
h_{\rm s}(r, H) \propto Nu(r, H).\\
q(r, H) \propto \Delta T_{\rm s}(r, H).
\end{eqnarray}
Therefore, the following equation relating the Nusselt number $Nu(r, H)$ to the temperature change $\Delta T_{\rm s}(r, H)$ is valid.
\begin{equation}
\Delta T_{\rm s}(r, H) = \alpha (T_{\rm s0} - T_{\rm a}(H))Nu(r, H),
\end{equation}
where $\alpha$ is also an unknown coefficient that needs to be estimated. Furthermore, we refer to \cite{a1} and assume that the temperature $T_{\rm a}(H)$ of the airflow changes as a function of the present distance $H$.
\begin{equation}
T_{\rm a}(H) = T_{\rm a0} + (T_{\rm e} - T_{\rm a0})(1 - e^{-gH}),
\end{equation}
where $T_{\rm a0}$ is the initial temperature of airflow, $T_{\rm e}$ is the ambient temperature, $g$ is an unknown coefficient that needs to be estimated.

\section{EXPERIMENT 1—EVALUATION OF AIRFLOW HEAT-TRANSFER MODEL}
\par To fit the proposed airflow heat-transfer model, we conducted an experiment measuring temperature changes.

\subsection{Experimental conditions and settings}
\begin{figure}[b]
\begin{center}
\includegraphics*[scale=0.26]{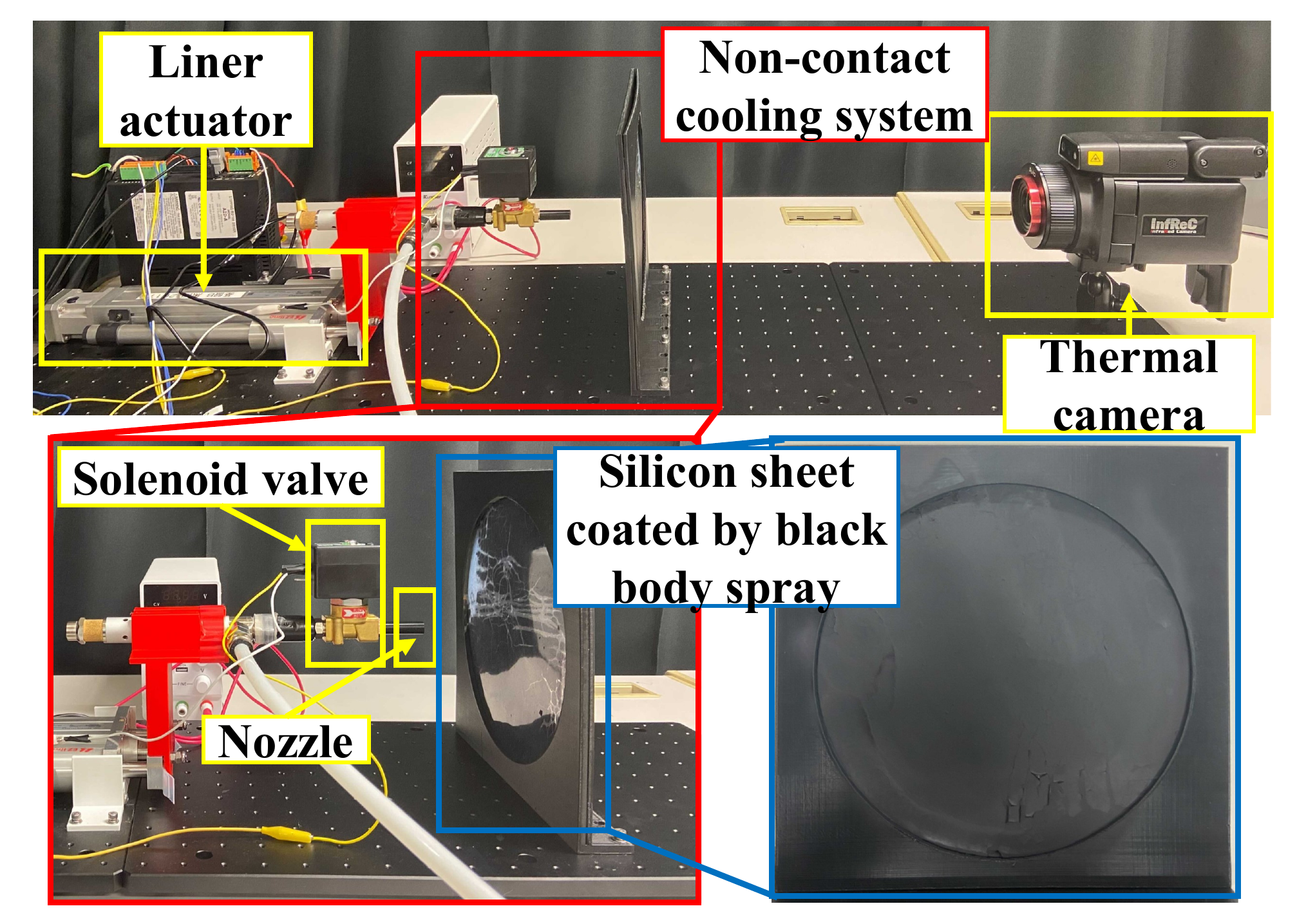}
\end{center}
\vspace*{-1mm}
\caption{Experimental settings for measuring using a silicone sheet.}
\label{fig:modelenv}
\end{figure}

As shown in Fig.~\ref{fig:modelenv}, we developed a measurement system to collect data for fitting. The system consists of a vortex tube (Tohin/AC-50), a proportional solenoid valve (Asco/Positive-flow-202), and a linear actuator (Oriental motor/AZM46AC). 
The vortex tube first generated low-temperature airflow. Then, the solenoid valve controlled the output volume flow rate of airflow and the linear actuator adjusted the presentation distance. The solenoid valve and the linear actuator were controlled by serial communication with a microcontroller (Arduino UNO). During the measurement, the room temperature $T_{\rm e}$ was set at 25~\({}^\circ\)C and the vortex tube was supplied with compressed air of 0.6 to 0.8~Mpa generated by a compressor (Hitachi/POD-0.75LES), thus the output air temperature $T_{\rm a0}$ was almost constant of 0~\({}^\circ\)C. The Prandtl number $Pr$ of air at 0~\({}^\circ\)C is 0.7~\cite{Pr}. We used a square silicone sheet with a thickness of 1~mm and side lengths of 200~mm as a phantom and coated it with blackbody paint (TASCO TA410KS with emissivity of 0.94). We set the presentation distance $H$ at 10, 20, 30, 40, and 50~mm, and the output volume flow rate at 25~L/min. As the nozzle inner diameter $d_{0}$, which represents the characteristic length of the airflow, is 6~mm, the Reynolds number $Re$ of air at 25~L/min is 5851\cite{Re}. We used a thermal camera (Avionics/InfReC R450, temperature resolution: $\leq 0.025~{}^\circ$C) to measure the temperature distribution of the silicone sheet after giving a cold stimulus for 4~s. Each condition was measured 5 times. To obtain accurate measurements, it was necessary to position the silicon sheet and the thermal camera facing each other and capture the thermal image from the backside.
Before each measurement, we placed the silicone sheet on a medium-temperature hot plate (NISSIN/NHP-M20) to initialize its temperature $T_{\rm s0}$ to almost 33~\({}^\circ\)C, which is about the general skin temperature. In actual, the silicone sheet has to be moved to the presentation site. Thus, its temperature might change.

\subsection{Results}
\par
\begin{figure}[htbp]
\begin{center}
\includegraphics*[scale=0.22]{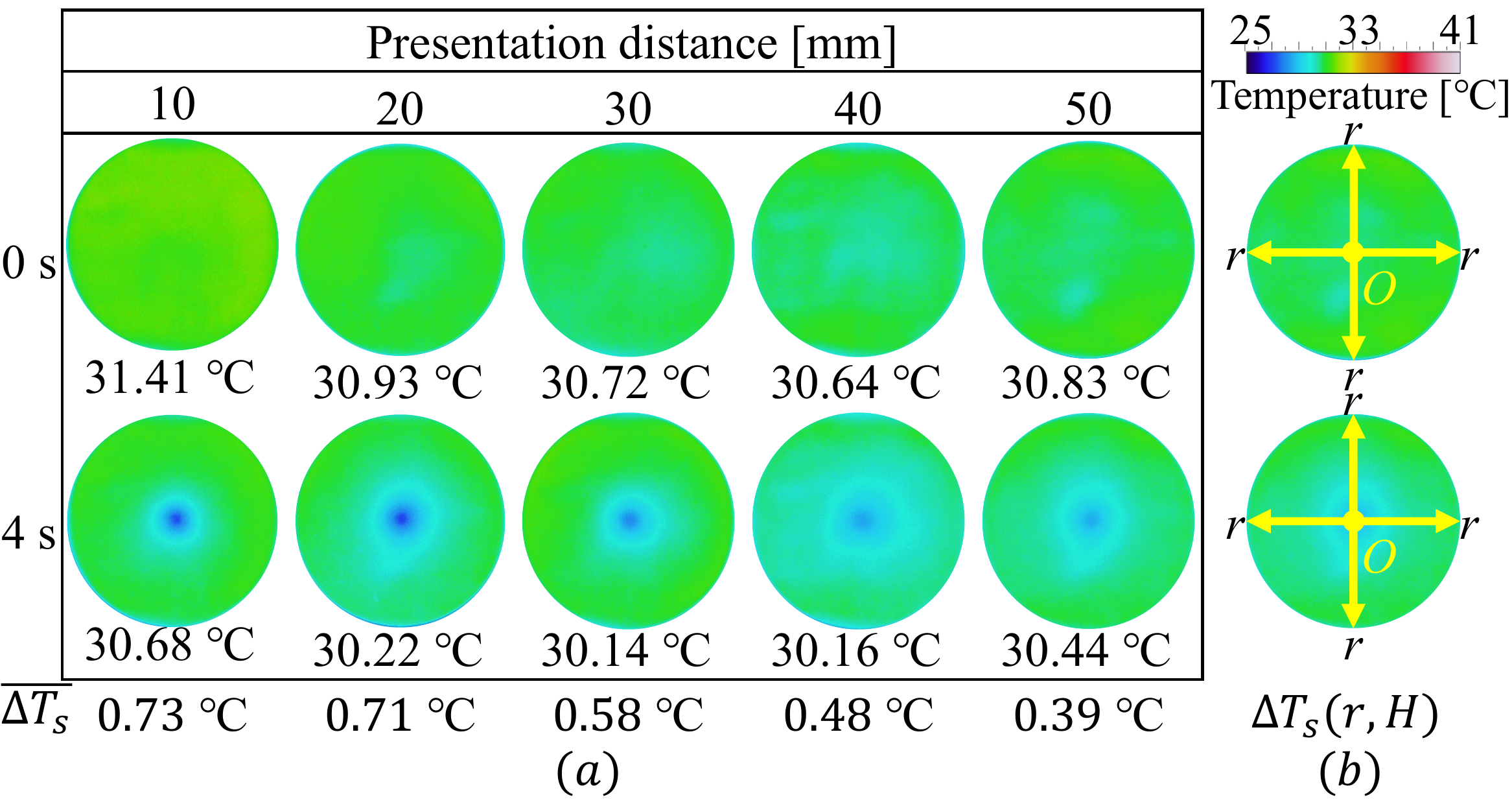}
\end{center}
\vspace*{-1mm}
\caption{Temperature distributions of the silicone sheet measured (a). An example of the temperature changes that were obtained for fitting the proposed model (b).}
\label{fig:thermal}
\begin{center}
\includegraphics*[scale=0.32]{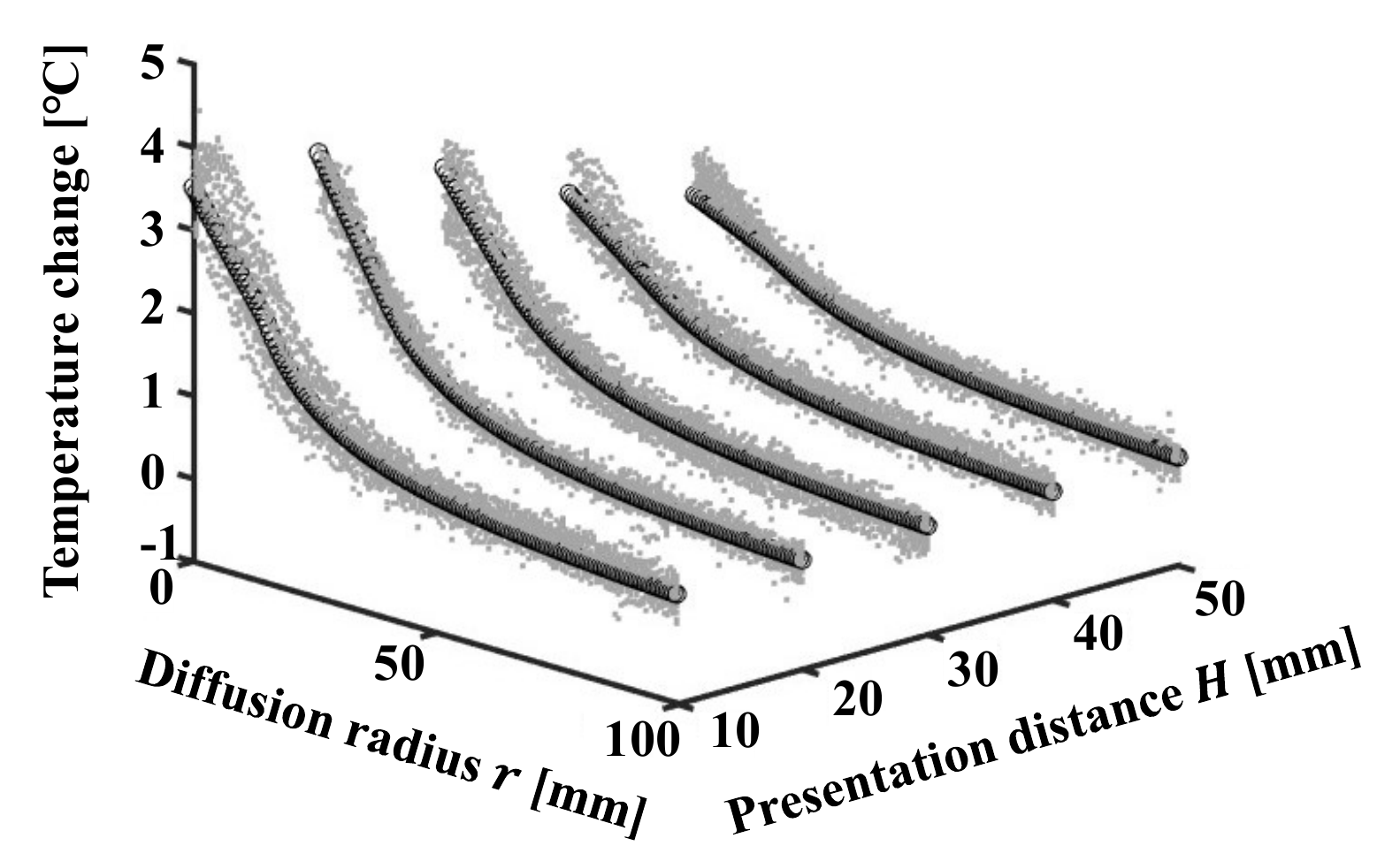}
\end{center}
\vspace*{-1mm}
\caption{Predicted results and measured temperature changes at different presentation distances.}
\label{fig:fit}
\end{figure}

 Fig.~\ref{fig:thermal}~(a) shows the measured temperature distributions of the silicone sheet, where each circle represents the temperature distribution on a circular area corresponding to a specific presentation distance. The results indicate that the temperature difference between 0~s and 4~s decreases as the presentation distance (10, 20, 30, 40, 50~mm) increases, with values of 0.73, 0.71, 0.58, 0.48, and 0.39~\({}^\circ\)C, respectively.
As shown in Fig.~\ref{fig:thermal}~(b), the temperature changes $\Delta T_{\rm s}(r, H)$ from center $O$ were obtained in four directions for fitting the proposed model. Using these temperature changes $\Delta T_{\rm s}(r, H)$, we estimated the coefficients for the proposed model using the iterative least-squares method, obtaining $\alpha=7.41$, $\beta=0.921$, $\gamma=141$, $a=7.87$, $b=0.145$, $c=9.33\times10^{-4}$, $f=2.63\times10^{-10}$, $g=1.72$ and $n=4.00$ with a coefficient of determination of 0.92 and a root mean-square error (RMSE) of 0.20~$^\circ$C for all measurements. Fig.~\ref{fig:fit} shows the predicted results and measurement temperature changes. The gray points represent the measured temperature changes and the black solid lines represent the predicted results. The results show that our airflow heat-transfer model can suitably fit temperature changes $\Delta T_{\rm s}(r, H)$ in different presentation distances $H$ and diffusion radius $r$. 
\section{EXPERIMENT 2—INVESTIGATION OF THE LOCAL COLD STIMULUS GROUP DISCRIMINATION THRESHOLD}
\par To investigate the LCSGDTs at different presentation distances, we conducted an experiment using the method of limits~\cite{we,pp}. As well, we discussed the relationship between the LCSGDTs and the heat-transfer capacity of low-temperature airflow.
\subsection{Participants}
\par For this experiment, 12 paid participants (aged 21--30 years, ave 23.5±2.53 years, 1 female and 11 males) were enrolled. Each participant received approximately USD 8 
in the form of an Amazon gift card as monetary compensation. The recruitment of participants and experimental procedures were approved by the Ethical Committee of the Department (Deleted for anonymity).
All participants provided written informed consent prior to taking part in the study.
\subsection{Experimental conditions and settings}
\par
\begin{figure}[t]
\begin{center}
\includegraphics*[scale=0.26]{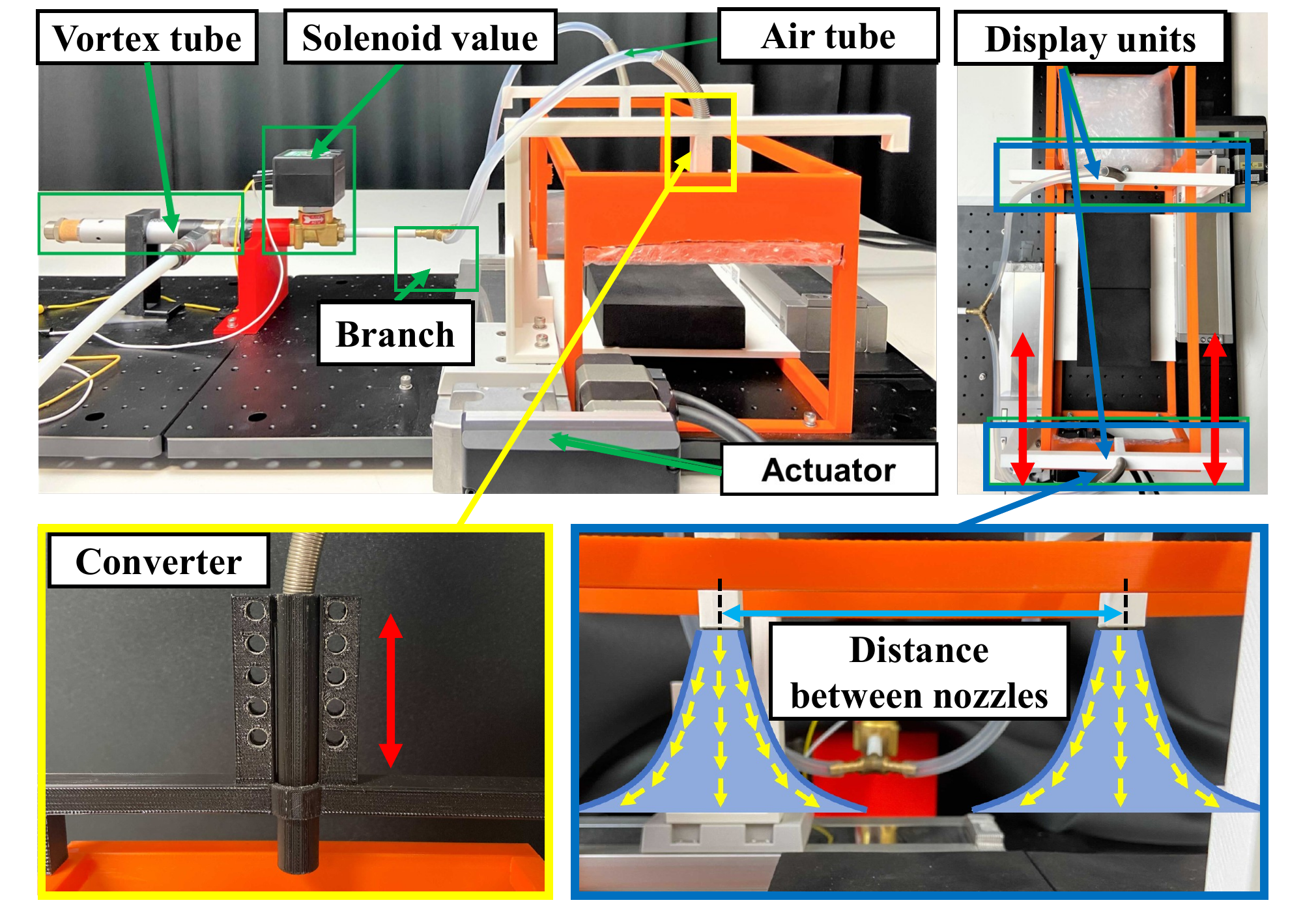}
\end{center}
\vspace*{-5mm}
\caption{System for investigating the local cold stimulus group discrimination threshold.}
\label{fig:sys}
\end{figure}

  As shown in Fig.~\ref{fig:sys}, we developed a LCSGDT measurement system. The system mainly consists of a vortex tube (Tohin/AC-50) that generates low-temperature airflow and a presentation component that divides the airflow into two equal portions. The presentation component was manufactured using a 3D printer (Raise3D/Pro2 with PLA filament) and is equipped with two linear actuators (Oriental motor/AZM46AC) to adjust the distance between the airflows. A proportional solenoid valve (Asco/Positive-flow-202) was used to control the output volume flow rate of airflow to 50~L/min. Accordingly, the output volume flow rate of divided airflow is 25~L/min. The two linear actuators and the solenoid valve are controlled by serial communication with a microcontroller (Arduino/UNO). We measured the LCSGDTs at different presentation distances of 10, 20, 30, 40, 50~mm, using the method of limits. This is because, among the psychophysical measurement methods, the limit method is simple and requires less time. We measured each condition three times. In each measurement, the distance between the airflows was controlled to 50, 70, 90, 110, 130, 150, 170, 190, 210, 230~mm. The lower limit of 50 mm was determined empirically, and the upper limit of 230 mm was determined by considering the length of the forearm of the experiment participants.
\begin{figure}[t]
\begin{center}
\includegraphics*[scale=0.27]{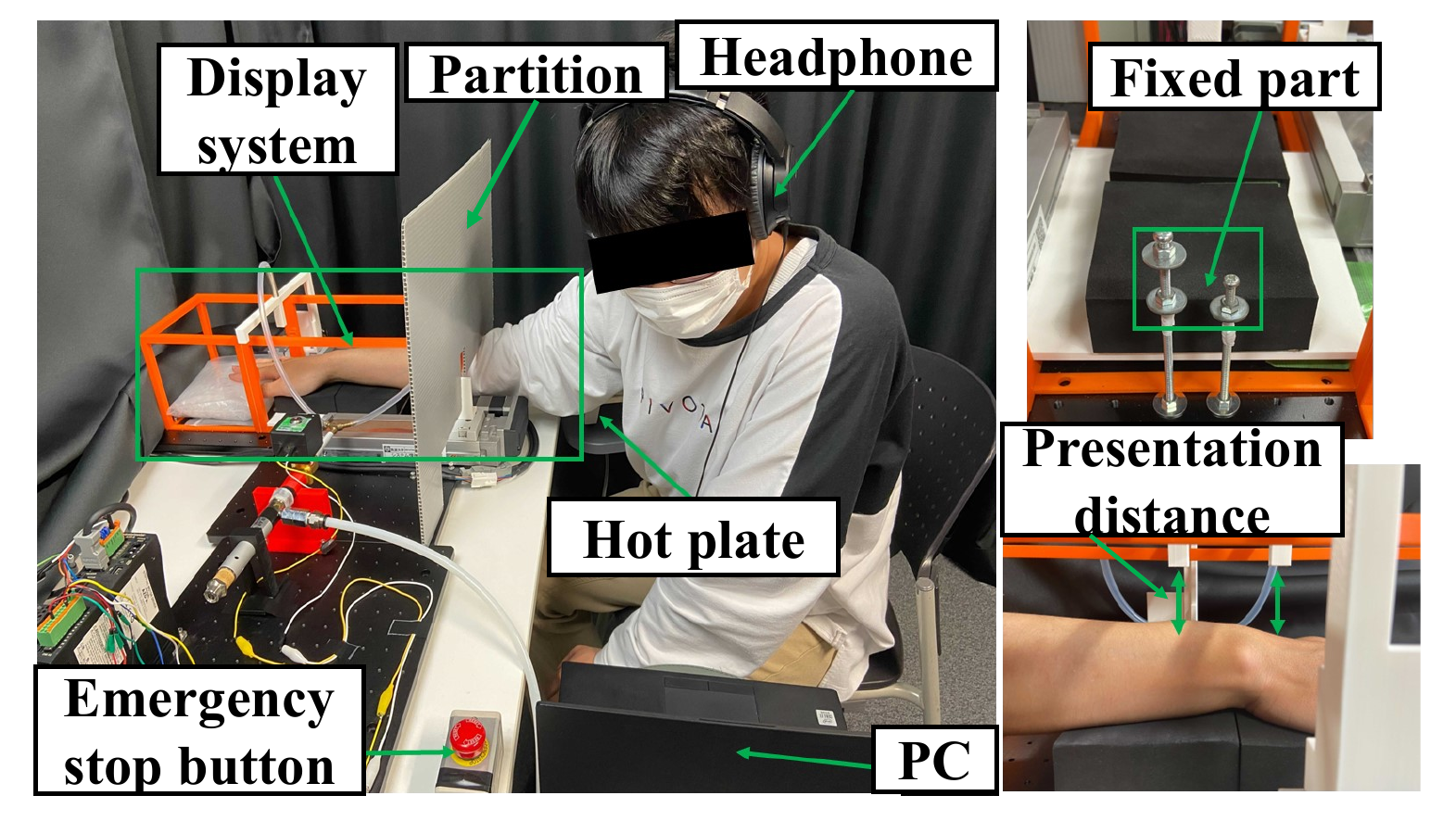}
\end{center}
\vspace*{-5mm}
\caption{Experimental settings for investigating the local cold stimulus group discrimination threshold.}
\label{fig:ex1}
\end{figure}
\par As shown in Fig.~\ref{fig:ex1}, during the experiment, we eliminated the influence of visual and auditory information by placing a partition between the LCSGDT measurement system and the experimental participants and asking participants to wear noise-cancelling headphones. Additionally, participants were required to fix their forearms by placing their fingers in the area designated for finger fixation on the middle finger. This was done to eliminate the influence of different placements of the forearms in each measurement. During the measurement, the room temperature was set at 25~\({}^\circ\)C using an air conditioner.
\par 
Before each measurement, participants first placed their outer forearms on a medium-temperature hot plate, which was set to the general skin temperature of 33~\({}^\circ\)C, for 30~s to initialize the skin temperature. Next, participants placed their outer forearms into the LCSGDT measurement system. To ensure that the forearms were positioned correctly, we checked that their middle fingers were inserted into the fixation area. After completing the previous step, we began the ascending series, gradually increasing the distance between airflows from 50 to 230~mm. At each distance, we gave cold stimuli for 4~s and asked participants to indicate if they felt two different local cold stimuli by responding with either "yes" or "no" displayed on the PC monitor. The ascending series was completed once the participants responded with a "yes" by pressing the keyboard. Next, we instructed participants to place their outer forearms on a medium-temperature hot plate to reset their skin temperature, and began the descending series by gradually decreasing the distance between airflows from 230 to 50~mm. The descending series was completed when participants responded with a "no" by pressing the keyboard. We calculated the LCSGDT as the average distance at which participants answered "yes" in the ascending series and "no" in the descending series.

\subsection{Results}
\begin{figure}[t]
\begin{center}
\includegraphics*[scale=0.29]{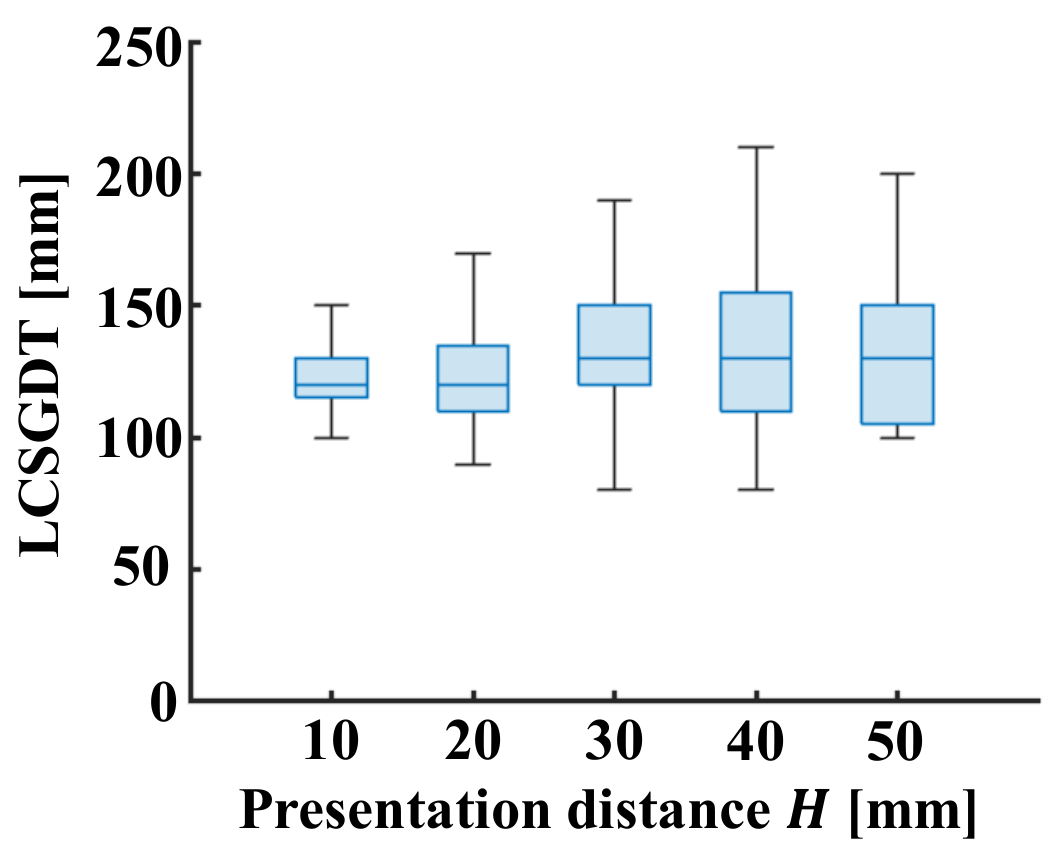}
\end{center}
\vspace*{-5mm}
\caption{The results of LCSGDT at different presentation distances.}
\label{fig:disr}
\end{figure}

Fig.~\ref{fig:disr} shows the results of the LCSGDT at different presentation distances. When the presentation distance was 40~mm, the LCSGDT reached its maximum value of 136.1±4.7~mm, and when the presentation distance was 10~mm, the LCSGDT reached its minimum value of 125.6±2.9~mm. 
A friedman one-way repeated measure analysis of variance
was performed on these measurements because no normality was found, and the LCSGDT did not demonstrate statistically significant differences in factor presentation distance (df=4, p=0.07). 
The mean value of the LCSGDT was 131.4±1.9~mm. We calculated the Nusselt number $Nu(r, H)$ corresponding to the LCSGDT at each presentation distance. The calculation results are shown in Table~\ref{table:po}. The result shows that the Nusselt number was almost constant, 0.92±0.00. The results support our hypothesis that the LCSGDT relies on the heat-transfer capability of low-temperature airflow.

\begin{table}[t]
\caption{The Nusselt number corresponding to the LCSGDT at each presentation distance.}
\label{table:po}
\centering
\begin{tabular}{|c||c|c|c|c|c|}
\hline
\begin{tabular}{c}
\textbf{Presentation} \\\textbf{distance [mm]} 
\end{tabular}
& 10  &  20  & 30  &  40  &  50  \\
\hline
\begin{tabular}{c}
\textbf{Nusselt}\\ \textbf{number [-]} 
\end{tabular}
& 0.923  &  0.925  &  0.924 &  0.920&  0.919\\
\hline
\end{tabular}
\end{table}
\section{DISCUSSION}
\par In experiment 1, we used a silicone sheet instead of skin and measured the temperature change to fit the proposed airflow heat-transfer model. The coefficient of determination for fitting the results was 0.92 and the RMSE was 0.20~$^\circ$C. To use the proposed model, it is possible to calculate the Nusselt number $Nu(r,h)$ at any diffusion radius $r$ and presentation distance $H$. In the result analysis, we only used the measurement data obtained from the four directions shown in Fig.~\ref{fig:thermal}. Although it was sufficient to fit the model, in future analyses, additional directions should also be taken into consideration.
Furthermore, the current model does not take into account the thermal properties of the body in the tangential direction during heat transfer. This factor should be taken into account in future analyses.
\par In experiment 2, we investigated the LCSGDTs at different presentation distances, and confirmed that the Nusselt number corresponding to all the LCSGDTs was almost the same, 0.92. This indicates that the LCSGDTs could be determined using the airflow heat-transfer model. When the presentation distance $H$ is given, we can calculate $r$ satisfying $Nu(r, H)=0.92$, and twice this $r$ is the LCSGDT. On the other hand, the volume flow rate of airflow could potentially influence the LCSGDTs and cold sensations. This is because as the volume flow rate changes, it affects the heat-transfer capacity and the reach of the airflow, which in turn affects the intensity of the cold sensation. In the future, we will study how changes in the volume flow rate impact the LCSGDTs and cold sensations in detail. Additionally, the cutaneous sensation caused by the airflow may also affect the LCSGDTs, and we plan to investigate this further.
\par In this study, 11 of the 12 participants were male, with only 1 female participant. Thus, we plan to increase the number of female participants in future studies to investigate the differences in LCSGDT between males and females. Additionally, we only examined the outer forearms of the participants. In future studies, we will expand our analysis to other body parts, such as the neck and legs, with the aim of creating the sensation of being in a cold environment, like a snow-covered mountain, by providing minimal airflow to the necessary areas of the skin.

\section{CONCLUSION}
\par
\begin{figure}[htbp]
\begin{center}
\includegraphics*[scale=0.23]{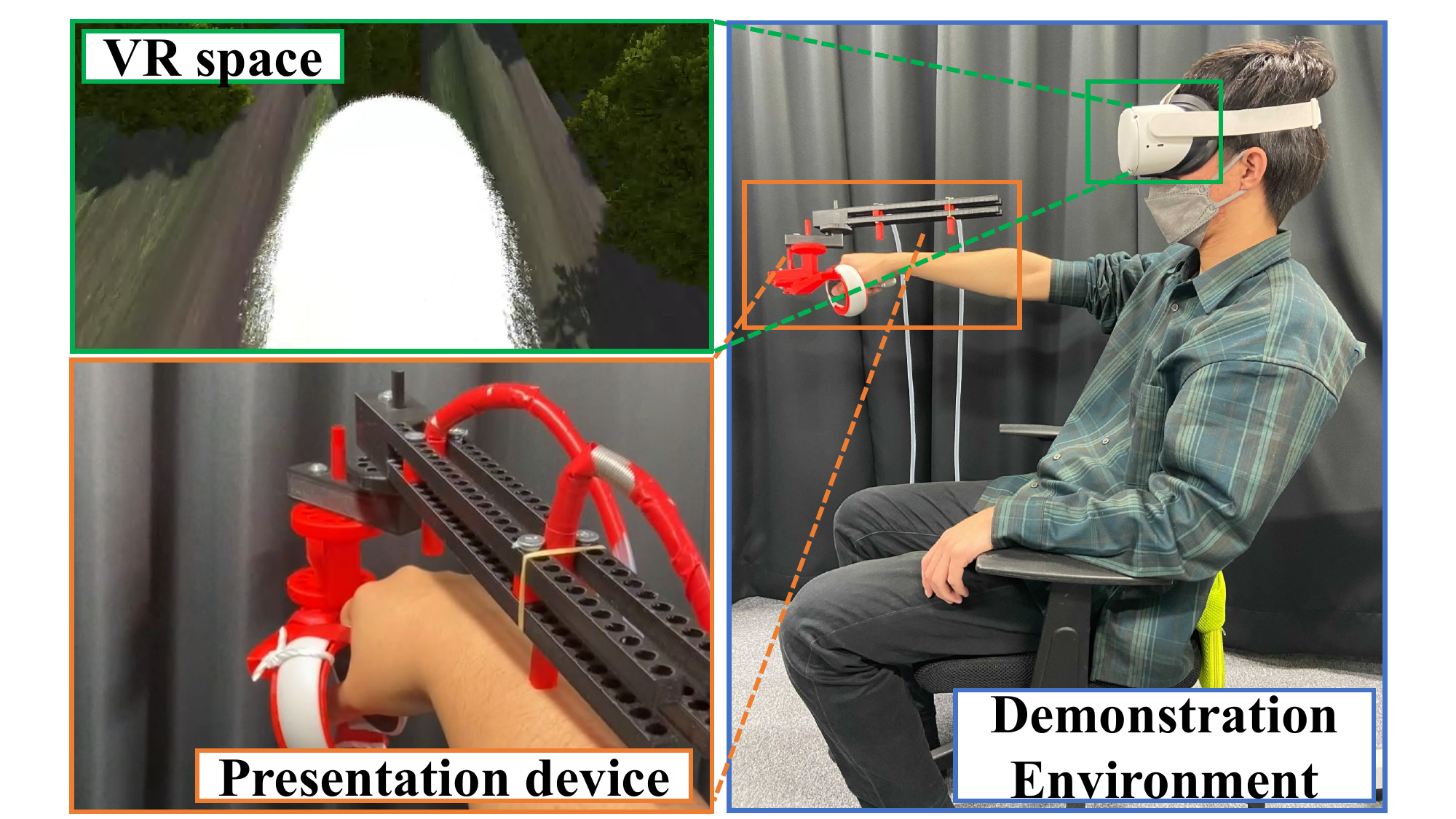}
\end{center}
\vspace*{-1mm}
\caption{A thermal display that can simulate a virtual cold space by controlling the distance between airflows within the LCSGDT.}
\label{fig:demo}
\end{figure}

 In this study, we proposed a method that realizes spatially continuous cold sensations based on low-temperature airflows. To this end, we considered the LCSGDT, which is the shortest distance between low-temperature airflows that are perceived as separate cold stimuli. Our hypothesis was that the LCSGDT is related to the heat-transfer capacity of low-temperature airflows. Therefore, we developed an airflow heat-transfer model and conducted an experiment to evaluate it. The experimental results showed that our model is capable of predicting the heat-transfer capacity of low-temperature airflow at different presentation distances. In subsequent experiments, we investigated the LCSGDT and its relationship with the heat-transfer capacity of low-temperature airflow. We found that the Nusselt number corresponding to all the LCSGDTs was almost the same, 0.92, which confirmed our hypothesis. 
By using our proposed model, it is possible to predict the LCSGDT at different present distances without prior investigation. Our future plans include the development of a thermal display that simulates a virtual cold space by controlling the distance between airflows within the LCSGDT, as shown in Fig.~\ref{fig:demo}.



\begin{thebibliography}{99}
\bibitem{warmAndCool} L. A. Jones, and H. Ho, ``Warm or cool, large or small? The challenge of thermal displays,'' IEEE Transactions on Haptics, vol. 1, no. 1, pp.~53--70, 2008.
\bibitem{m0} M. Nakajima, Y. Makino, and H. Shinoda. ``Remote cooling sensation presentation controlling mist in midair,'' in Proceedings of 2020 IEEE/SICE International Symposium on System Integration (SII), pp. 1238-1241, 2020.
\bibitem{m1}
M. Nakajima, K. Hasegawa, Y. Makino, and H. Shinoda, ``Spatiotemporal Pinpoint Cooling Sensation Produced by Ultrasound-Driven Mist Vaporization on Skin,'' IEEE Transactions on Haptics, 
vol.~14, no.~4, pp.~874--884, 2021.
\bibitem{a0} J. Xu, Y. Kuroda, S. Yoshimoto, and O. Oshiro. ``Non-contact cold thermal display by controlling low-temperature air flow generated with vortex tube,'' in Proceedings of 2019 IEEE World Haptics Conference (WHC), pp. 133-138, 2019.
\bibitem{a1} 
J. Xu, S. Yoshimoto, N. Ienaga, and Y. Kuroda, ``Intensity-Adjustable Non-contact Cold Sensation Presentation Based on the Vortex Effect,''
IEEE Transactions on Haptics, vol.~15, no.~3, pp.~592--602,
2022.
\bibitem{p0}
R. L. Peiris, W. Peng, Z. Chen, L. Chan, and K. Minamizawa, ``Thermovr: Exploring integrated thermal haptic feedback with
head mounted displays,'' In Proceedings of the 2017 CHI Conference on Human Factors in Computing Systems, ACM, pp.~5452--5456, 2017.
\bibitem{p1}
K. Ito, Y. Ban, and S. Warisawa, ``Coldness Presentation Depending on Motion to Enhance the Sense of Presence in a Virtual Underwater Experience,'' IEEE Access, vol. ~10, pp.~23463--23476, 2022.
\bibitem{w1}
S. G\"{u}nther, F. M\"{u}ller, D.  Sch\"{o}n, O. Elmoghazy, M. M\"{u}hlh\"{a}user, and M. Schmitz, ``Therminator: Understanding the Interdependency of 
Visual and On-Body Thermal Feedback in Virtual Reality,'' In Proceedings of the 2020 CHI Conference on Human Factors in Computing Systems, pp.~1--14, 2020. 
\bibitem{first}P. Han, Y. Chean, K. Lee, H. Wang, C. Hsieh, J. Hsiao, C. Chou, and Y. Hung, ``Haptic around: multiple tactile sensations for immersive environment and interaction in virtual reality,'' 
In Proceedings of the 24th ACM symposium on virtual reality software and technology, pp.~1--10, 2018.
\bibitem{AoEs}
P. Han, C. Hsieh, Y. Chen, J. Hsiao, K. Lee, S. Ko, K. Chen, C. Chou, and Y. Hung, ``AoEs: enhancing teleportation experience in immersive environment with mid-air haptics,''
In ACM SIGGRAPH Emerging Technologies, 2 pages, 2017. 
\bibitem{vt}
G. J. Ranque, ``Experiments on Expansion in Vortex with Simultaneous Exhaust of Hot and Cold Air,''
Le Journal De Physique et le Radium (Paris),
vol.~4, pp.~112--114, 1933.
\bibitem{Nu}
G. F. Hewitt, G. L. Shires, and T. R. Bott, ``Process Heat Transfer,'' CRC Press, 1994.
\bibitem{a} 
H. Martine, ``Heat and Mass Transfer Between Impinging Gas jets and Solid Surfaces,'' Advances in Heat Transfer, vol.~13, pp.~1--60, 1970.
\bibitem{T}
N. Gerrett, Y. Ouzzahra, and G. Havenith, ``Distribution of
Skin Thermal Sensitivity,'' Springer International Publishing,
pp.~1--17, 2016.
\bibitem{Pr}
C. Y. Li, and S. V. Garimella, ``Prandtl-number effects and generalized correlations for confined and submerged jet impingement,'' International Journal of Heat and Mass Transfer, pp.~3471--3480, 2001.
\bibitem{Re}
J. Happel, and H. Brenner, ``Low Reynolds number hydrodynamics: with special applications to particulate media,'' Springer Science \& Business Media, vol.~1, 1983.
\bibitem{we}
G. Werner, ``The study of sensation in physiology: psychophysical and neurophysiologic correlations,'' Medical physiology, 1974.
\bibitem{pp} 
G. A. Gescheider, ``Psychophysics: The Fundamentals,'' Psychology Press, 1997.



\end{thebibliography}
\end{document}